\documentclass[two column]{jpsj3} 
\usepackage{amsmath,amssymb}
\usepackage{txfonts}
\usepackage{graphicx}
\usepackage{color}

\title{$^{59}$Co-Nuclear Quadrupole Resonance and Nuclear Magnetic Resonance studies on YCoGe \\
  --- Comparison between YCoGe and UCoGe ---}

\author{
Kosuke \text{Karube}$^{1,}$\thanks{karube@scphys.kyoto-u.ac.jp}, 
Taisuke Hattori$^1$, 
Yoshihiko Ihara$^{1,+}$, 
Yusuke Nakai$^{1,2}$, 
Kenji Ishida$^{1,2,}$\thanks{kishida@scphys.kyoto-u.ac.jp}, 
Nobuyuki Tamura$^3$, 
Kazuhiko Deguchi$^3$, 
Noriaki K. Sato$^3$,
and Hisatomo Harima$^4$} 

\inst{$^1$Department of Physics, Graduate School of Science, Kyoto University, Kyoto 606-8502, Japan\\
$^2$Transformative Research Project on Iron Pnicides (TRIP), Japan Science and Technology Agency (JST),\\
Chiyoda, Tokyo 102-0075, Japan\\
$^3$Department of Physics, Graduate School of Science, Nagoya University, Nagoya 464-8602, Japan\\
$^4$Department of Physics, Graduate School of Science, Kobe University, Kobe 657-8501, Japan   
}

\abst{We have performed $^{59}$Co-nuclear quadrupole resonance (NQR) and nuclear magnetic resonance (NMR) studies on YCoGe, which is a reference compound of ferromagnetic superconductor UCoGe, in order to investigate the magnetic properties at the Co site. Magnetic and superconducting transitions were not observed down to 0.3 K, but a conventional metallic behavior was found in YCoGe, although its crystal structure is similar to that of UCoGe. From the comparison between experimental results of two compounds, the ferromagnetism and superconductivity observed in UCoGe originate from the U-5$f$ electrons.}

\kword{YCoGe, UCoGe, NMR, nuclear quadrupole resonance}

\begin{document}
\maketitle

\maketitle
\section{Introduction} 
The discovery of superconductivity in ferromagnet UGe$_2$ under pressure gave a great impact for researchers studying superconductivity\cite{letter_UGe2}, because ferromagnetism and superconductivity have been thought to be mutually exclusive. 
In 2007, a similar superconductivity was discovered in ferromagnet UCoGe at ambient pressure by Huy \textit{et al}\cite{letter_UCoGe}. Ferromagnetic (FM) and superconducting (SC) transition temperatures 
($T_\mathrm{Curie}$ and $T_\mathrm{SC}$) of UCoGe were reported to be 3 and 0.8 K, respectively\cite{letter_UCoGe_highHc2}.  
In addition, $\mu$SR and $^{59}$Co-NQR measurements suggest that ferromagnetism and superconductivity microscopically coexist, and that the SC gap is formed in the FM region\cite{letter_UCoGe_muSR,letter_UCoGe_single}.
In UCoGe, it has been believed that the U-5$f$ electrons are responsible for ferromagnetism and superconductivity from the analogy of UGe$_2$. However, there is a possibility that ferromagnetism in UCoGe originates from Co-3$d$, not from U-5$f$ electrons, because it is well known that Co-3$d$ electrons give rise to magnetism in some Co compounds, and there is a report indicating that the Co-3$d$ electrons contribute to the magnetism in UCoGe in a high field (12 T)\cite{letter_UCoGe_polarizedND}.  

\begin{figure}[tb]
\begin{center}
\includegraphics[scale=0.15]{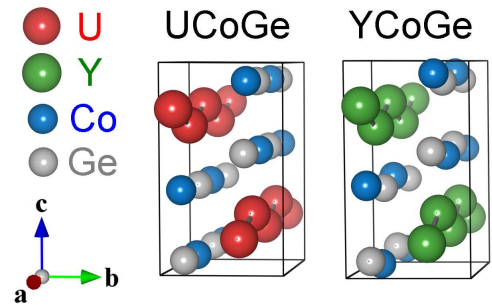}
\caption{(Color online) TiNiSi-type orthorhombic crystal structures of UCoGe and YCoGe. The difference between the two structures is the alignment of Co-Ge.}
\label{crystal_structure_YCoGe&UCoGe}
\end{center}
\end{figure}

To clarify the role of Co-3$d$ electrons to ferromagnetism in UCoGe, we point out that YCoGe would be a good reference compound for UCoGe, because Y has no $f$ electrons and YCoGe has the similar TiNiSi-type crystal structure and lattice constants\cite{letter_YCoGe_structure} to UCoGe\cite{letter_UCoGe_structure}, as shown in Fig.~\ref{crystal_structure_YCoGe&UCoGe}. 
In addition, the band calculation suggests that the contribution of Co-3$d$ electrons to the density of states 
is quite similar both in UCoGe and YCoGe\cite{Harima}.
Although the crystal structure of YCoGe studied by X-ray diffraction analysis was reported in the literature
\cite{letter_YCoGe_structure}, its physical properties have not been reported so far.
In this paper, we report $^{59}$Co-NMR and nuclear quadrupole resonance (NQR) results as well as electrical resistivity and specific-heat results in YCoGe measured down to 0.3 K. The results strongly suggest that the ferromagnetism and superconductivity observed in UCoGe originate from the U-5$f$ electrons.     

\section{Experimental Procedure} 
Polycrystalline and single-crystal samples of YCoGe were prepared by arc-melting and Czochralski-pulling methods with a tetra-arc furnace, respectively. From X-ray diffraction measurements, small peaks assigned to impurity phases were observed in a polycrystalline sample, but were not in a single-crystal sample. 
This implies that the quality of the single-crystal sample is higher than that of the polycrystalline sample.
Tiny single crystals were crushed into fine powder and packed in a sample case made from a straw of 5 mm diameter. 
The powder was mixed with GE varnish, stirred, and fixed to a random orientation in order to avoid a preferential orientation under magnetic fields.

\begin{figure}[htb]
\begin{center}
\includegraphics[scale=0.76]{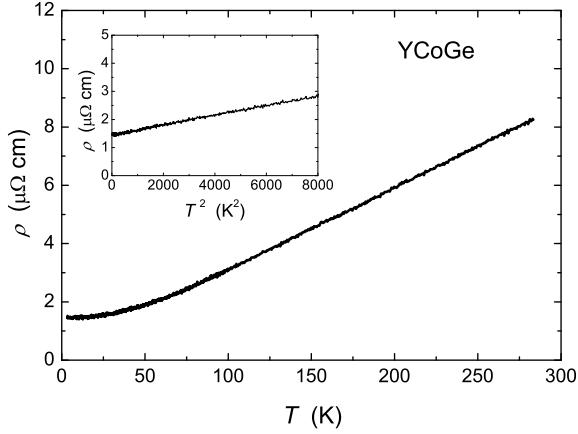}
\caption{Temperature dependence of the resistivity of YCoGe down to 3.4 K. 
The inset displays the resistivity as a function of temperature squared.}
\label{YCoGe_resistivity}
\end{center}
\end{figure}
\begin{figure}[htb]
\begin{center}
\includegraphics[scale=0.8]{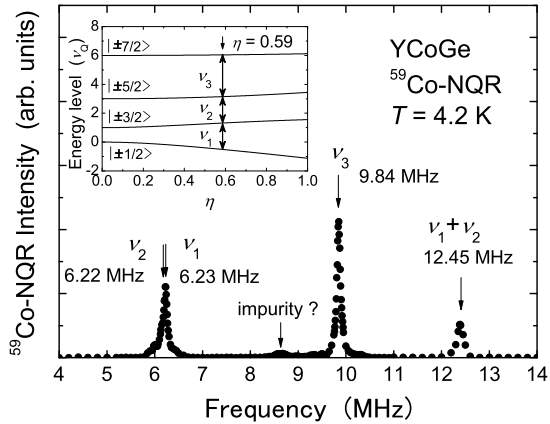}
\caption{$^{59}$Co-NQR spectrum of YCoGe at 4.2 K. Intensities are normalized at each peak.
The inset exhibits nuclear spin energies for $I$ = 7/2 (in non-axial EFG) as a function of $\eta$. 
The analysis yields $\nu_1$ = 6.23 MHz, $\nu_2$ = 6.22 MHz and $\nu_3$ = 9.84 MHz, and 
the quadrupole parameters $\nu_Q$ = 3.40 MHz and $\eta$ = 0.59.}
\label{NQRspectrum_YCoGe}
\end{center}
\end{figure}
\section{Experimental Results}
Before showing NMR and NQR results, bulk properties on YCoGe are overviewed. 
As shown in Fig.~\ref{YCoGe_resistivity}, the resistivity in the single-crystal sample exhibits a typical metallic behavior without magnetic and superconducting anomalies. The inset indicates that low-temperature resistivity is proportional to $T^2$ below 100 K, which is characteristic of the Fermi-liquid state.
The specific heat ($C$) was measured on the single-crystal sample down to 0.3 K. Below 10 K, the experimental data $C$ was well fitted by $C = \gamma T + \beta T^3$, and the electronic ($\gamma$) and phonon ($\beta$) coefficients were evaluated as $\gamma$ = 6.6 (mJ/mol$\cdot$K$^2$) and $\beta$ = 0.134 (mJ/mol$\cdot$K$^4$), respectively. 
 
We performed $^{59}$Co nuclear quadrupole resonance (NQR) on YCoGe, and searched NQR signals over a wide frequency range from 3 to 20 MHz.
Three sharp peaks, displayed in Fig.~\ref{NQRspectrum_YCoGe}, were observed, whose frequencies are 6.23, 9.84, and 12.45 MHz at 4.2 K.
The NQR Hamiltonian is provided as,
\begin{eqnarray}
\mathcal{H}_{Q}=\frac{\hbar\nu_{Q}}{6}\left\{ 3(I_{z}^{2}-I^2)+\frac{\eta}{2}(I_{+}^{2}+I_{-}^{2}) \right\}, 
\end{eqnarray}
where $\nu_Q$ is the frequency along the principal axis of the electric field gradient (EFG) 
and $\eta$ is the asymmetry parameter, defined as $\eta\equiv (V_{xx}-V_{yy})/V_{zz}$. 
Here, $V_{ij}$ is the component of the EFG tensor.
When $^{59}$Co with nuclear spin $I=7/2$ is in the presence of EFG, the degenerate eight nuclear-spin states 
$\left|m\right> (m = 7/2, \cdots, -7/2)$ are split into four energy levels by electric quadrupole interaction, yielding three resonance frequencies, $\nu_1$, $\nu_2$ and $\nu_3$, as shown in the inset of Fig.~\ref{NQRspectrum_YCoGe}.
One may consider that the signals observed at 6.23 and 12.45 MHz arise from the $\nu_1$ and $\nu_2$ ($\nu_2$ and $\nu_3$) signals with $\eta$ = 0, but this possibility is excluded since the $\nu_3$ ($\nu_1$) signal is not observed at 18.7 MHz (3.1 MHz). 
The observed three frequencies cannot be interpreted by the $\nu_1$, $\nu_2$ and $\nu_3$ transitions; alternatively, the observed NQR peaks are assigned as follows. $\nu_1$ and $\nu_2$ are almost overlapped and observed at 6.23 and 6.22 MHz, and $\nu_3$ is observed at 9.84 MHz. The 12.45 MHz peak is assigned to $\nu_1+\nu_2$ =  12.45 MHz, which is caused by hybrid states due to nonzero $\eta$. From the assignment, $\nu_Q$ = 3.40 MHz and $\eta$ = 0.59 are derived, and the wave functions of the four energy levels are expressed, as shown in Table \ref{table1}.
\begin{table}
\caption{Wave functions of four energy levels of the Co nuclear spin in YCoGe.}
\begin{tabular}{c|cccc}
 &$|\pm7/2 \rangle $ & $|\pm5/2 \rangle $ &  $|\pm3/2 \rangle $ & $|\pm1/2 \rangle$ \\  \hline
$\Psi_{\pm7/2}$ & -0.996 & -0.003 & -0.090 & -0.012\\
$\Psi_{\pm5/2}$ & -0.012 &   0.972 &  0.075 &   0.221\\
$\Psi_{\pm3/2}$ &   0.085 &   0.166 & -0.886 & -0.425\\
$\Psi_{\pm1/2}$ &   0.031 &  -0.164 & -0.449 &  0.878
\end{tabular}
\label{table1}
\end{table}
In this case, the resonance peak corresponding to $\nu_2+\nu_3 = 16.06$ MHz is expected, but the transition probability between $\Psi_{\pm7/2}$ and $\Psi_{\pm3/2}$ is one magnitude smaller than that between $\Psi_{\pm5/2}$ and $\Psi_{\pm1/2}$; thus, we could not observe signals around 16 MHz.
A tiny signal was observed at around 8.7 MHz, which is thought to arise from impurity phases.

\begin{figure}[tb]
\begin{center}
\includegraphics[scale=0.8]{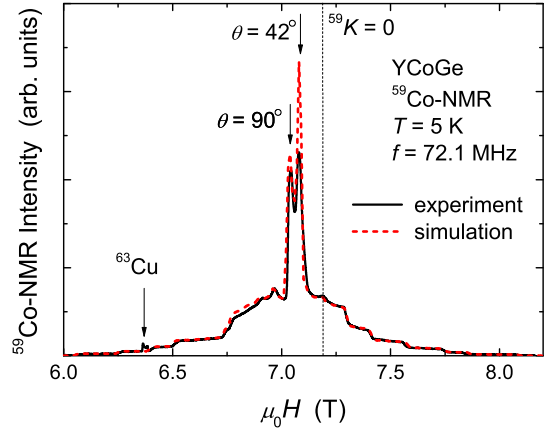}
\caption{(Color online) Field-swept $^{59}$Co-NMR spectrum at 5 K. The black solid line shows the experimental spectrum 
and the red dotted line shows the random powder simulation result.
$^{59}K$ = 0 shows the field where the isotropic Knight shift of $^{59}$Co is zero.
The two central peaks marked by arrows are derived from $\theta$ = 90 and 42$^\circ$. 
The small peak arising from $^{63}$Cu in the coil is also 
marked by an arrow.}
\label{NMRspectrum_YCoGe}
\end{center}
\end{figure}
To confirm the validity of the NQR-parameter identification, we performed $^{59}$Co-NMR measurements. The field-swept $^{59}$Co-NMR spectrum, shown in Fig.~\ref{NMRspectrum_YCoGe}, was obtained from the powdered single-crystal sample at 5 K. The red dotted line in Fig.~\ref{NMRspectrum_YCoGe} shows the result of the simulation\cite{MetalicShift}, in which the principal axis of EFG is randomly distributed against the external field, and  $\nu_Q$ = 3.40 MHz and $\eta$ = 0.59, which are determined by the above mentioned NQR measurement, and the isotropic Knight shift $K_\mathrm{iso}=1.88\%$ and the anisotropic Knight shift $K_\mathrm{aniso}=-0.28\%$ are used. The simulation result is in good agreement with the observed NMR spectrum, indicating that the NQR parameters can also be used to interpret the NMR spectrum. 
The Knight-shift values in YCoGe are comparable to those in UCoGe at higher temperatures. The NQR and Knight-shift values of YCoGe and UCoGe are summarized in Table \ref{table2}. 
\begin{table}
\caption{NQR and Knight-shift parameters as well as an electronic coefficient in the specific-heat measurements in YCoGe and UCoGe.}
\begin{tabular}{c|cc}
\hline 
& YCoGe & UCoGe \\ \hline
 $\nu_Q$(exp.) [MHz] & 3.40 & 2.85 \\
 $\eta$(exp.) &  0.59 & 0.52 \\
$\nu_Q$(band) [MHz] & 3.91 & 3.11  \\
$\eta$(band) & 0.33 & 0.47 \\ 
$K_{\rm iso}$ [\%] &  1.88 (5 K) & 1.58 (200 K)\\
$K_{\rm aniso}$ [\%] & -0.28 (5 K) & 0.02 (200 K) \\ 
$\gamma$ [mJ /mol K$^2$] & 6.6 & 65 \\     
\hline
\end{tabular}
\label{table2}
\end{table}

\begin{figure}[tb]
\begin{center}
\includegraphics[scale=0.8]{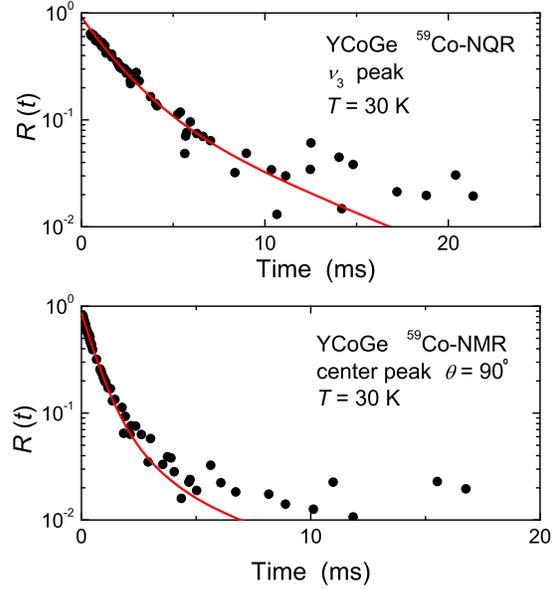}
\caption{(Color online) Recovery curves $R(t)$ of the nuclear magnetization $m(t)$ at a time $t$ after a saturation pulse with the theoretical curves for evaluating $1/T_1$. (a) $R(t)$ measured at $\nu_3$ in Fig.~\ref{NQRspectrum_YCoGe}, and (b) $R(t)$ measured at the $\theta=90^{\circ}$ peak in the NMR spectrum of Fig.~\ref{NMRspectrum_YCoGe} (see in text). } 
\label{recovery}
\end{center}
\end{figure}

The nuclear spin-lattice relaxation rate $1/T_1$ of Co was measured with the NMR and NQR methods in order to investigate the anisotropy of magnetic fluctuations. In general, $1/T_1$ is determined with the fluctuations of the hyperfine fields perpendicular to the quantum axis. $1/T_1$ was measured at the $\nu_3$ peak in the $^{59}$Co-NQR spectrum, whose $1/T_1$ detects magnetic fluctuations along the $b$- and $c$-axis directions, since the EFG principal axis is considered to be almost parallel to the $a$-axis from the analogy of UCoGe\cite{letter_UCoGe_Ising}. $1/T_1$ was derived by fitting the recovery curves $R(t) = 1-m(t)/m(\infty)$ with the following theoretical NQR recovery curve for $\nu_3$\cite{letter_NQR_recovery},
\begin{eqnarray}
R(t)&=&0.163\exp\left(\frac{-2.74t}{T_1}\right)+0.675\exp\left(\frac{-9.22t}{T_1}\right)\nonumber\\
     &&+0.162\exp\left(\frac{-17.9t}{T_1}\right).
\end{eqnarray}
Here, $m(t)$ is the nuclear magnetization measured at a time $t$ after a saturation pulse. 
$1/T_1$ was also measured at the NMR central peaks corresponding to  $\theta$ = 90 and 42$^{\circ}$ shown with arrows in Fig.~\ref{NMRspectrum_YCoGe}, where $\theta$ is the angle between the external field and the EFG principal axis. 
The $1/T_1$ measured at the $\theta$ = 90$^{\circ}$ peak detects the magnetic fluctuations including in the $a$-axis direction, since the EFG principal axis is almost $a$-axis.
The $1/T_1$ measured at the NMR peaks is derived by fitting with the theoretical recovery curve for the central transition\cite{letter_NMR_recovery}. 
The experimental $R(t)$ and theoretical curves for the NQR and NMR measurements are shown in Fig.~\ref{recovery}, and $1/T_1$ was derived from the reasonable results of fitting.  
\begin{figure}[tb]
\begin{center}
\includegraphics[scale=0.8]{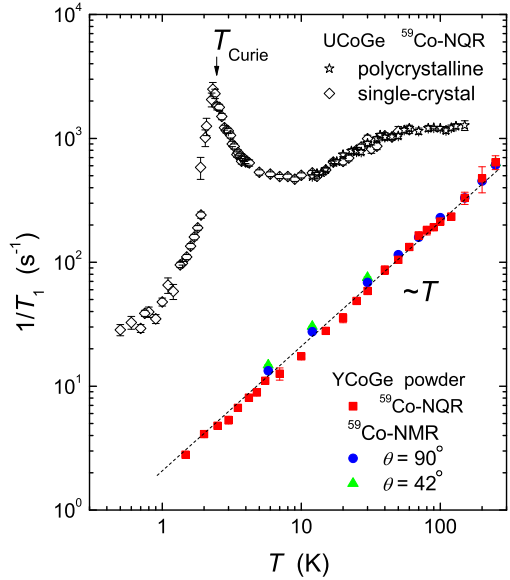}
\caption{(Color online) Temperature dependence of $^{59}$Co-NQR $1/T_1$ (red closed square) and  
$^{59}$Co-NMR $1/T_1$ for $\theta$ = 90$^{\circ}$ (blue closed circle) and 
$\theta$ = 42$^{\circ}$ (green closed triangle) in YCoGe, 
compared with $^{59}$Co-NQR $1/T_1$ in polycrystalline (open star) and single-crystal (open diamond) 
UCoGe\cite{letter_UCoGe_single}. 
The ferromagnetic transition temperature in UCoGe marked by $T_\mathrm{Curie}$ was reported to be 2.5 K.} 
\label{1/T1vsTinYCoGe&UCoGe}
\end{center}
\end{figure}
The $1/T_1$ results are shown in Fig.~\ref{1/T1vsTinYCoGe&UCoGe}.
We found that $1/T_1$ is isotropic and proportional to a temperature above 1.5 K. 
The isotropic $T_1T$ = const. (so-called ``Korringa'') behavior indicates that YCoGe is in a conventional metal state without notable magnetic fluctuations. The $1/T_1$ result is consistent with the resistivity and specific-heat results.
By assuming the Korringa relationship  between the constant $T_1 T$ and the spin part of the Knight shift ($K_\mathrm{s}$), $K_\mathrm{s}$ is described by
\begin{eqnarray}
K_\mathrm{s}=\sqrt{\frac{1}{T_1T}\frac{\hbar}{4\pi k_\mathrm{B}}}\left(\frac{\gamma_\mathrm{e}}{\gamma_\mathrm{n}}\right), 
\end{eqnarray}
 where $\gamma_\mathrm{e}$ and $\gamma_\mathrm{n}$ are gyromagnetic ratios with respect to electrons and nuclei, respectively. 
Using the experimental values of $1/T_1 T$ (= 2.03 s$^{-1}$K$^{-1}$), $K_\mathrm{s}$ is estimated as $K_\mathrm{s}$ = 0.31\%, which is much smaller than the experimental value of the Knight shift ($K_{\rm iso} \sim$ 1.88\%); thus, the orbital part of the Knight shift $K_\mathrm{orb}\sim$ 1.57\% is derived, since the experimental value of the Knight shift is the sum of $K_\mathrm{s}$ and $K_\mathrm{orb}$. 
The experimental value of $K_\mathrm{orb}$ is a reasonable value of $K_\mathrm{orb}$ reported in several Co compounds such as a Co metal ($K_{\rm orb}=1.7$ \%)\cite{letter_Co} and nonmagnetic NaCoO$_2$ ($K_{\rm orb}=1.9$ \%)\cite{letter_NaCoO2}.

\section{Discussion}
Here, we compare our results in YCoGe with those in UCoGe\cite{letter_UCoGe_single}. 
First, YCoGe has a similar $^{59}$Co-NQR spectrum to UCoGe, although $\nu_1$ and $\nu_2$ are overlapped, indicative of the low symmetry of the crystal structure. As shown in Table \ref{table2}, quadrupole parameters in YCoGe are slightly greater than those in UCoGe.
The difference can be interpreted by the difference in the Co-Ge alignment along the $a$-axis. The Co-Ge alignment is almost straight along the $a$-axis in UCoGe, {whereas it is zigzag in YCoGe (see Fig~\ref{crystal_structure_YCoGe&UCoGe}). Therefore, the local symmetry of Co atoms in YCoGe is lower than that in UCoGe, providing larger quadrupole parameters. However, the asymmetric parameter $\eta$ in YCoGe calculated by band calculation is smaller than that in UCoGe, which is inconsistent with the experimental observation (see Table \ref{table2}). This contradiction seems to be caused by the electronic state of the Y site in the band calculation.    

As for magnetic fluctuations, the $1/T_1$ in UCoGe is much greater than that in YCoGe in the entire temperature range and displays a prominent large peak due to FM ordering at around $T_{\mathrm{Curie}}$ $\sim$ 2.5 K, as seen in Fig.~\ref{1/T1vsTinYCoGe&UCoGe}. In addition, recent direction-dependent Co-NMR measurements in single-crystal UCoGe revealed that the Knight shift and $1/T_1$ show a significant anisotropic behavior, indicating the presence of the Ising-type FM fluctuations along the magnetic easy axis ($c$-axis) at low temperatures below 10 K\cite{letter_UCoGe_Ising}. It is also shown that such characteristic FM fluctuations are intimately related to the unconventional superconductivity in UCoGe\cite{letter_UCoGe_Ising}. Such magnetic fluctuations were not observed at all in YCoGe. However, it should be noted that the experimental values of $K$ and $1/T_1$ at 230 K in UCoGe are isotropic and comparable to those in YCoGe\cite{letter_UCoGe_Ising}.
This implies that the high-temperature electronic state in UCoGe is similar to that in YCoGe, since the hybridization between conduction electrons and U - 5$f$ local moments is weak at high temperatures and the electronic state in UCoGe is governed by the conduction electron; thus, YCoGe is a good reference compound for understanding the electronic state without U-5$f$ electrons. Therefore, we conclude that the strong Ising-type FM fluctuations accompanied by the weak FM ordering and unconventional superconductivity as well as the heavy-fermion character are ascribed to the U-5$f$ electrons. However, the field-induced magnetism observed in UCoGe\cite{letter_UCoGe_polarizedND}, which is related to a 5$f$-3$d$ hybridization, is an important issue in the understanding of the high magnetic state in UCoGe. High-field $^{59}$Co-NMR measurements are expected to provide crucial information on this issue.

\section{Conclusion}
From $^{59}$Co-NQR/NMR measurements, we determined NQR parameters in YCoGe, which are similar to those in UCoGe. 
The temperature dependence of $1/T_1$ shows that YCoGe is in a conventional metallic state without notable magnetic fluctuations down to 1.5 K, which is consistent with resistivity and specific-heat measurements.
From the comparison between the experimental results in YCoGe and those in UCoGe, we conclude that U-5$f$ electrons simultaneously carry ferromagnetism and unconventional superconductivity in UCoGe.

\section*{Acknowledgments}
The authors thank D. Aoki and J. Flouquet for valuable discussions, and D. C. Peets, S. Yonezawa, and Y. Maeno for experimental support and valuable discussions. This work was partially supported by Kyoto Univ. LTM Centre, the ``Heavy Electrons'' Grant-in-Aid for Scientific Research on Innovative Areas  (No. 20102006, No. 21102510, and No. 20102008) from the Ministry of Education, Culture, Sports, Science, and Technology (MEXT) of Japan, a Grant-in-Aid for the Global COE Program ``The Next Generation of Physics, Spun from Universality and Emergence'' from MEXT of Japan, and a Grant-in-Aid for Scientific Research from the Japan Society for Promotion of Science (JSPS), KAKENHI (S) (No. 20224015).

\vspace{1cm}
\noindent
+present address: Department of Physics, Faculty of Science, Hokkaido University, Nishi 8, Kita 10, Kita-ku, Sapporo 060-0810, Japan


\begin{thebibliography}{99}
\bibitem{letter_UGe2} S. S. Saxena, P. Agarwal, K. Ahilan, F. M. Grosche, R. K. W. Haselwimmer, M. J. Steiner, 
E. Pugh, I. R. Walker, S. R. Julian, P. Monthoux, G. G. Lonzarich, A. Huxley, I. Sheikin, D. Braithwaite, 
and J. Flouquet: Nature (London) \textbf{406} (2000) 587.

\bibitem{letter_UCoGe} N. T. Huy, A. Gasparini, D. E. de Nijs, Y. Huang, J. C. P. Klaasse, T. Gortenmulder, A. de Visser, A. Hamann, T. G\"orlach, and H. v. L\"ohneysen: Phys. Rev. Lett. \textbf{99} (2007) 067006.

\bibitem{letter_UCoGe_highHc2} N. T. Huy, D. E. de Nijs, Y. K. Huang, and A. de Visser: Phys. Rev. Lett. \textbf{100} (2008) 077002.

\bibitem{letter_UCoGe_muSR} A. de Visser, N. T. Huy, A. Gasparini, D. E. de Nijs, D. Andreica, C. Baines, and A. Amato: Phys. Rev. Lett. \textbf{102} (2009) 167003.

\bibitem{letter_UCoGe_single} T. Ohta, T. Hattori, K. Ishida. Y. Nakai, E. Osaki, K. Deguchi, 
N. K. Sato, and I. Satoh: J. Phys. Soc. Jpn. \textbf{79} (2010) 023707.

\bibitem{letter_UCoGe_polarizedND} K. Proke, A. de Visser, Y. K. Huang, B. Fak, and E. Ressouche: Phys. Rev. B \textbf{81} (2010) 180407(R).

\bibitem{letter_YCoGe_structure} A. E. Dwight, P. P. Vaishnava, C. W. Kimball, and J. L. Matykiewicz: 
J. Less-Common. Met. \textbf{119} (1986) 319.

\bibitem{letter_UCoGe_structure} F. Canepa, P. Manfrinetti, M. Pani, and A. Palenzona: J. Alloys Comd. \textbf{234} (1996) 225.

\bibitem{Harima} H. Harima: private communication.

\bibitem{MetalicShift} G. C. Carter, L. H. Bennett, and D. J. Kahan: \textit{Metallic Shift in NMR} (Pergamon, New York, 1977).

\bibitem{letter_UCoGe_Ising} Y. Ihara, T. Hattori, K. Ishida, Y. Nakai, E. Osaki, K. Deguchi, 
N. K. Sato, and I. Satoh: Phys. Rev. Lett. \textbf{105} (2010) 206403.

\bibitem{letter_NQR_recovery} J. Chepin and J. H. Ross, Jr: J. Phys. Condens. Matter. \textbf{3} (1991) 8103.

\bibitem{letter_NMR_recovery} A. Narath: Phys. Rev. \textbf{162} (1967) 320.

\bibitem{letter_Co} U. El-Hanany and  W. W. Warren, Jr.: Bull. Am. Phys. Soc. \textbf{19} (1974) 202.

\bibitem{letter_NaCoO2} G. Lang, J. Bobroff, H. Alloul, P. Mendels, N. Blanchard, and G. Collin: Phys. Rev. B \textbf{72} (2005) 094404.

\end{thebibliography}
\end{document}